\documentclass{article}
\usepackage[T1]{fontenc}
\usepackage[latin9]{inputenc}
\usepackage{amstext}
\usepackage{graphicx}
\usepackage{amsmath}
\usepackage{color}
\usepackage{authblk}
\usepackage{algorithm,algorithmic}
\usepackage{multirow}
\usepackage{float}
\usepackage[caption = false]{subfig}
\newcommand{\abo}{[\alpha/\beta]_i}
\newcommand{\abt}{[\alpha/\beta]_T}

\makeatother

\begin{document}

\title{Minimizing Metastatic Risk in Radiotherapy Fractionation Schedules}
\author[1]{Hamidreza Badri}
\author[2]{Jagdish Ramakrishnan}
\author[1]{Kevin Leder}
\affil[1]{Industrial and Systems Engineering, University of Minnesota, Minneapolis, MN, USA}
\affil[2]{Wisconsin Institute for Discovery, University of Wisconsin, Madison, WI, USA}

\date{\today}

\maketitle
\begin{abstract}
Metastasis is the process by which cells from a primary tumor disperse and form new tumors at distant anatomical locations. The treatment and prevention of metastatic cancer remains an extremely challenging problem. This work introduces a novel biologically motivated objective function to the radiation optimization community that takes into account metastatic risk instead of the status of the primary tumor. In this work, we consider the problem of developing fractionated irradiation schedules that minimize production of metastatic cancer cells while keeping normal tissue damage below an acceptable level. A dynamic programming framework is utilized to determine the optimal fractionation scheme. We evaluated our approach on a breast cancer case using the heart and the lung as organs-at-risk (OAR). For small tumor $\alpha/\beta$ values, hypo-fractionated schedules were optimal, which is consistent with standard models. However, for relatively larger $\alpha/\beta$ values, we found the type of schedule depended on various parameters such as the time when metastatic risk was evaluated, the $\alpha/\beta$ values of the OARs, and the normal tissue sparing factors. Interestingly, in contrast to standard models, hypo-fractionated and semi-hypo-fractionated schedules (large initial doses with doses tapering off with time) were suggested even with large tumor $\alpha$/$\beta$ values. Numerical results indicate potential for significant reduction in metastatic risk.
\end{abstract}

\section{Introduction}
Most solid tumors eventually establish colonies in distant anatomical locations; when these colonies become clinically detectable, they are called macrometastasis. While often there is a large burden from primary tumors, it is in fact metastatic disease that is responsible for most cancer fatalities \cite{GuMa06,weinberg2013biology}.

The creation of macrometastasis requires the successful completion of a sequence of difficult steps. First, cancer cells must gain access to the general circulation system via the process of intravasation. Next, the cells must survive in the inhospitable environment of the circulatory system. Following this, the tumor cells must exit the circulatory system (extravasation) at a distant site and initiate micrometastsis (clinically undetectable population of tumor cells at a distant anatomical site). Lastly, the micrometastsis must develop the ability to successfully proliferate in the distant site and grow into clinically identifiable macrometastasis. The completion of these steps is very difficult and only a small fraction of tumor cells are able to achieve this \cite{MePu06}. However, due to the vast number of cells in most primary tumors, metastasis commonly occurs in later stage solid tumors. 

There has been significant mathematical research in the design of optimal anti-cancer therapies. This has included studies on optimal chemotherapy, radiotherapy, and more recently targeted therapies and immunotherapy (\cite{Swan1990,KaKo10,AiBe10,Foo_PlosCB_Res,CaPi07}). Since we are interested in radiotherapy we will focus on previous work in this field. The vast majority of modeling of radiotherapy response is based on the linear-quadratic model (LQ) which says that tissue response is governed by the parameters $\alpha$ and $\beta$ (see e.g., \cite{HallGiaccia}). Specifically, following a single exposure to $d$ Gray of radiation, the surviving fraction of viable cells is given by $\exp(-\alpha*d-\beta*d^2)$. An important question in this field is to decide on the optimal temporal distribution of a given amount of radiation, i.e., how to kill the most tumor cells while inflicting the least amount of normal tissue damage. This is commonly referred to as the `optimal fractionation problem.' Two possible solutions to this problem are hyper-fractionated  and hypo-fractionated schedules. In hyper-fractionated schedules, small fraction sizes are delivered over a large number of treatment days, while in hypo-fractionated schedules, large fraction sizes are delivered over a small number of treatment days. If we minimize primary tumor cell population at the conclusion of treatment, it has been seen (\cite{MiTaDa12} and \cite{Badrimath}) that whether hyper or hypo-fractionation is preferable depends on the radiation sensitivity parameters of the normal and cancerous tissue. However we will observe in Section 4 of this manuscript that when designing optimal treatments with the goal of minimizing metastatic production, hypo-fractionation is preferable for many parameter choices, and hyper-fractionation is only preferable sometimes when the $\alpha/\beta$ value of the tumor is large.

There have been a substantial number of works looking at optimal fractionation.
The work \cite{WeCoWu} considers dynamic design of fractionation schedules with incomplete repair, repopulation and reoxygenation. A more recent work \cite{BeBrPaSi} considers the optimization problem associated with finding fractionation schedules under an LQ model with incomplete repair and exponential repopulation. The authors theoretically establish the benefits of hypo-fractionation in the setting of a low $\alpha/\beta$ value of the tumor. Brenner and Hall \cite{BrHa} utilized the LQ model in combination with the Lea-Catcheside function (a generalization of the LQ model that is useful at higher doses or prolonged doses) to conclude that due to its slow response to radiation, prostate cancer can be treated equally effectively by either uniform radiation scheduling or hypo-fractionation (which has fewer side effects). Unkelbach et al. \cite{UCS2013} studied the interdependence between optimal spatial dose distribution and creation of fractionation schedules. Another work \cite{BRT2013} utilized a dynamic programming approach to study the problem of optimal fractionation schedules in the presence of various repopulation curves. An important property common to all of these works is that they utilize an objective function that seeks to minimize final primary tumor population size in some sense. While this can be an important objective, in most cancers, it is ultimately metastatic disease that proves fatal. Therefore, in this work, we study optimal fractionation schedules when using an objective function that seeks to minimize the total production of metastatic cells.

The understanding of the metastatic process and how to respond to it has been greatly aided by the mathematical modeling community (for an overview of this contribution see the recent review paper \cite{ScGe13}). In an interesting work \cite{Iwata2000}, Iwata et al. developed a set of differential equations governing the population dynamics of the metastatic population. A compelling work is the paper by Thames et al. \cite{ThBuSm99} where they developed a mathematical model of the metastatic process to calculate risk from metastatic disease due to delay in surgery. Hanin and Korosteleva \cite{HaKo10} used a stochastic model to address questions such as: (1) how early do metastasis events occur, (2) how does extirpation of the primary affect evolution of the metastasis, and (3) how long are metastasis latent? Haeno and Michor \cite{HaMi10} developed a multitype branching process model to study metastasis and in particular the probability of metastasis being present at diagnosis. In a follow up work \cite{HaGo12}, they used a mathematical model to study metastasis data in recently deceased pancreatic cancer patients. In a recent work \cite{DiCaGa13}, Diego et al. used an ODE model to study the relationship between primary and metastatic cancer sites, and in particular, makes predictions about the clinical course of the disease based on the parameter space of their ODE model.

The remainder of the paper is organized as follows. In section \ref{sec:MetsProd}, we discuss a model for metastasis production and how it can be used to develop a function that reflects metastatic risk. Next, in section \ref{sec:Opt}, we describe the optimization model and solution approach. Finally, in section \ref{sec:num}, we present numerical results in the setting of breast cancer.

\section{A New Objective Function}
\label{sec:MetsProd}
We start by assuming that the total population of primary tumor cells at time $t$ is given by the function $X_t$. Note we will assume throughout this work that the population of cells is large enough that we can treat the population as a deterministic function. We then assume that each tumor cell initiates a successful macrometastasis at rate $\nu>0$ per unit of time. This is similar to the modeling approach taken in \cite{HaGo12} where they were able to fit their model to metastasis data from patients.
If we are interested in the time horizon $[0,T]$ then our total rate of production of successful macrometastasis on that interval is
$$
R_1(T)=\nu\int_0^TX_tdt.
$$
In particular $N(T)$, the number of successful metastasis established in the time interval in $[0,T]$, is a Poisson random variable with mean $R_1(T)$ and thus the probability of metastasis occurring is $P(N(T)>0)=1-\exp\left(-R_1(T)\right)$. Therefore, in order to minimize $P(N(T)>0)$, it suffices to minimize $R_1(T)$.

In the rate $R_1(T)$, we assume that every cell is capable of metastasis. In the geometry of the actual tumor, it might be the case that only those cells on the surface of the tumor are capable of metastasis, or only those cells in close proximity to a blood vessel are capable. Therefore, we consider the generalization 
$$
R_\xi(T)=\nu\int_0^T\left(X_t\right)^\xi dt
$$
for $0<\xi\leq 1$. This is similar to the model of metastasis creation in \cite{Iwata2000}; there they refer to the parameter $\xi$ as the fractal dimension of the blood vessels infiltrating the tumor,  If we assume that the tumor is three dimensional with a two dimensional surface and that all cells on the surface are equally capable of metastasis then we can take $\xi=2/3$. However if we assume that only a small fraction of cells on the surface are capable of metastasis we could take e.g., $\xi=1/3$. Notice that in order to minimize $R_\xi(T)$, we do not need to know the parameter $\nu$, which is difficult to measure. 

Note that we are using a rather simplistic model for the metastasis production in that we assume at most only two rates of metastasis for the primary tumor cells. In reality, it is likely that the rate of metastasis for a given cell will be a complex function of its position, migratory potential, and its oxidative state. However, given the lack of data available, we found it preferable to work with this relatively simplistic model that does not require the knowledge of any intricate parameters. In addition, since the primary goal of this work is to introduce a novel objective function, we feel that adding further biological details can be saved for further exploration. Lastly and importantly, variants of this relatively simplistic model have been matched to clinical metastasis data \cite{HaGo12}.

\section{Optimization Model and Approach}
\label{sec:Opt}

Our goal is to determine an optimal radiotherapy fractionation scheme
that minimizes the probability that the primary tumor volume metastasizes.
Given the discussion from the previous section, this equates to minimizing $R_\xi(T)$; however, for simplicity, we will use an approximate objective that uses a summation rather than an integral during therapy. Let the time horizon $T$ consists of treatment days and a potentially long period of no treatment at the end of which metastatic risk is evaluated. Suppose a radiation dose $d_t$ is delivered at time instant $t$, for $t = 1, \ldots, N_0$, then we choose to minimize $$\sum_{t=1}^{N_0} (X_{t}^{-})^\xi+\int_{N_0}^{T} (X_t)^\xi \approx R_\xi(T)/\nu$$ where $X_{t}^{-}$ is the number of cells immediately before the delivery of dose $d_{t}$. 

During the course of the treatment, we assume exponential tumor growth with a time lag. Beyond that, we use the Gomp-ex model for tumor growth because we are evaluating long-term metastatic risk and the exponential model will give unreasonably large values over such a long time period. The Gomp-ex law of growth assumes that the cellular population expands through the exponential law \cite{Wheldon} initially, when there is no competition for resources. However, there is a critical size threshold $C$ such that for $X>C$ the growth follows the Gompertz Law. Thus, we have
\begin{equation}
\label{eq:LinearQuadraticExp}
X_{t}^- = X_0 \exp \left(-(\alpha_T \sum_{i=1}^{t-1}d_i + \beta_T \sum_{i=1}^{t-1}d_i^2) + \frac{\ln 2}{\tau_d}(t-T_k)^+ \right), 
\end{equation}
for $t=1,\dots,N_0 $ and
\begin{equation}
\label{eq:LinearQuadraticExp1}
X_{t} =\begin{cases}X_0\exp\left( -(\alpha_T \sum_{i=1}^{N_0}d_i + \beta_T \sum_{i=1}^{N_0}d_i^2)+\frac{\ln 2}{\tau_d}(t-T_k)^+\right) ,& X_t\le C\\
K\exp\left( \log\left( \frac{C}{K}\right)e^{-\delta(t-T_c)} \right) ,& X_t> C\end{cases}
\end{equation}
for $t\in(N_0,T]$ where $\alpha_T$ and $\beta_T$ are tumor tissue sensitivity parameters, $\tau_d$ is the tumor doubling time in units of days, $T_k$ is the tumor kick-off time, $X_0$ is the tumor size at the starting of treatment, $K$ is the carrying capacity and $\delta$ is a constant related to the proliferative ability of the cells. The expression $(t-T_k)^+$ is defined
as $\max(0,t-T_k)$. $T_c$ is the time when the tumor population reaches size $C$ and can be computed using following equation
$$T_c=\frac{\tau_d}{\ln 2}\left( \ln\left( \frac{C}{X_0}\right) +\alpha_T\sum_{i=1}^{N_0}d_i+\beta_T\sum_{i=1}^{N_0}d_i^2\right)+T_k.$$
We also use the concept of biological effective dose (BED) to constrain the side-effects in the OAR around the tumor. We assume that a dose $d$ results in a homogeneous dose $\gamma_i d$ in the $i$th OAR, where $\gamma_i$ is the sparing factor; for the heterogeneous case, it is possible to use the approach in \cite{UCS2013}. The BED in the $i${th} OAR is defined by
\begin{equation}
\text{BED}_i = \sum_{t=1}^{N_0} \gamma_i d_t \left( 1 + \frac{\gamma_i d_t}{\abo} \right),
\end{equation}
where $\abo$ is an OAR tissue sensitivity parameter. The BED can be derived from the LQ model and is used to quantify fractionation effects in a clinical setting. Note that we can also have multiple BED constraints on a single OAR, e.g., to model early and late effects.

The optimization problem of interest is 
\small
\begin{equation}\label{opt}
\underset{{d_t\geq0}}{\text{minimize}}\quad\left(X_0 \right)^\xi \sum_{t=1}^{N_0} \exp \left(-\xi\left( \alpha_T \sum_{i=1}^{t-1}d_i + \beta_T \sum_{i=1}^{t-1}d_i^2-\frac{\ln 2}{\tau_d}(t-T_k)^+\right)\right)+\int_{N_0}^{T}(X_t)^\xi,
\end{equation}
s.t.
$$\quad\sum_{t=1}^{N_0} \gamma_i d_t \left( 1 + \frac{\gamma_i d_t}{\abo}\right) \leq c_i,\ \ i=1,\dots,M $$
\normalsize
where $c_i$ is a constant that specifies an upper bound to the $i${th} BED in the OAR and $M$ is the total number of OARs in the vicinity of tumor. Note that we do not work directly with the quantity $R_\xi$ during therapy but instead with its approximation $\hat{R}_\xi=\sum_{t=1}^{N_0}\left(X_t^-\right)^\xi$. Here, $\hat{R}_\xi$ is an upper bound for $R_\xi$ and is a good approximation for $R_\xi$ if the impact of the exponential growth term in (\ref{eq:LinearQuadraticExp}) is relatively small compared to the dose fraction terms, which is typically the case for most disease sites.

Formulation \eqref{opt} is a nonconvex quadratically constrained problem. Such problems are computationally difficult to solve in general. However, we can use a dynamic programming (DP) approach with only two states to solve this deterministic problem, similar to the work in \cite{BRT2013}. The states of the system are $U_t$, the cumulative dose, and $V_t$, the cumulative dose squared, delivered to the tumor immediately after time $t$. We have
$$U_{t}=U_{t-1}+d_{t}$$
$$V_{t}=V_{t-1}+d_{t}^2$$
Now we can write the DP algorithm (forward recursion) as
\tiny
$$J_{t}(U_{t},V_{t})=\begin{cases}\min_{d_{t}\ge0}[X_0^\xi e^{-\xi\left( \alpha_T (U_{t-1}+d_{t}) + \beta_T (V_{t-1}+d_{t}^2)-\frac{\ln 2}{\tau_d}(t-T_k)^+\right)}+J_{t-1}(U_{t-1},V_{t-1})] ,& t\le N_0-1\\
\min_{d_{t}\ge0}[X_0^\xi e^{-\xi\left( \alpha_T (U_{t-1}+d_{t}) + \beta_T (V_{t-1}+d_{t}^2)-\frac{\ln 2}{\tau_d}(t-T_k)^+\right)}+J_{t-1}(U_{t-1},V_{t-1})]+\int_{N_0}^{T}(X_t)^\xi ,& t=N_0\end{cases}$$
\normalsize
with $(U_0,V_0)=(0,0)$, $J_{0}(0,0)=0$. We set the function $J_t(U_t,V_t)$ to be $\infty$ if 
$$\gamma_i U_t+\frac{\gamma_i^2}{\abo} V_t>c_i$$
for any $i=1,\dots,M$. Since there are only two state variables, we can solve our optimization problem by discretizing the states and using this DP algorithm.

\section{Numerical Results}
\label{sec:num}

We solve the optimization problem based on the radiobiological parameters for breast cancer.  We consider two different normal tissues \cite{Joiner2009}, heart and lung tissue. For each normal tissue, we define the maximal toxicity  
\begin{equation}\label{bed}
c_i=\gamma_iD_{i}+\frac{\gamma_i^2}{\abo}\frac{D_i^2}{N_i}
\end{equation}
where $D_i$ and $N_i$ are tissue specific parameters and define the maximum total dose $D_i$ delivered in $N_i$ fractions for each OAR. A standard fractionated treatment is to deliver $50$ Gy to the tumor with $2$ Gy fractions \cite{stand1}. The tolerance BED values ($c_i$) for various normal tissues were computed based on the standard scheme. Hence all BED in OAR associated with optimal schedules obtained in this section are less than or equal to their corresponding BED in standard schedule, i.e. $D_i=50$ Gy and $N_i=25$ for $i=1,2$. All radiobiological parameter values used are listed in Table \ref{tabledata} along with their sources. 

\begin{table}[ht!]
	\begin{center}
    	\begin{tabular}{ | l | l | l | l | }
    	\hline
    	Structure & Parameters & Values  & Source \\ \hline
    	\multirow{8}{*}{Breast tumor} & $\alpha_T$ & $0.080$ Gy$^{-1}$ &  \cite{Qi2011}  \\ 
	    & $\beta_T$ & $0.028$ Gy$^{-2}$ &  \cite{Qi2011} \\ 
	    & $\tau_d$ & $14$ days &  \cite{Qi2011}  \\
	    & $T_k$ & $21$ days &  \cite{Tortorelli2013}  \\
	    & $X_0$ & $0.75\times10^{9}$ cells &  \cite{Monte2009}  \\
	    & $C$ & $4.8\times10^{9}$ cells &  \cite{Norton1998}  \\
	    & $K$ & $3.1\times10^{12}$ cells & \cite{Norton1998}  \\
	    & $\delta$ & $1.83\times 10^{-3}$ day$^{-1}$&  \cite{Norton1998} \\ \hline
	    \multirow{4}{*}{Lung normal tissue} & $\alpha/\beta$ & $5$ Gy &  \cite{Joiner2009} \\ 
	    & $\gamma$ & $8.30\%$ &  \cite{Das98} \\ 
	    & $c$ & $4.29$ Gy &  eq. \eqref{bed} \\ \hline
	    \multirow{4}{*}{Heart normal tissue} & $\alpha/\beta$ & $3$ Gy &  \cite{Joiner2009} \\ 
	    & $\gamma$ & $2.04\%$ &  \cite{Das98} \\ 
	    & $c$ & $1.03$ Gy &  eq. \eqref{bed} \\ \hline
    	\end{tabular}
    	\caption {Breast tumor and normal tissues parameters }
    	\label{tabledata}
	\end{center}
\end{table}

For other parameters in our model, unless stated otherwise, we set $T = 5$ years, $\xi = \frac{1}{3}$ and $N_0 = 25$ days.  For numerical implementation of the DP algorithm, we discretize the total dose and dose squared $(U_t,V_t)$. We used 2500 points to discretize each state variable for every time instant. When evaluating the cost-to-go function $J_{t}(U_t,V_t)$ for values in between discretization points, we use bilinear interpolation. The allowed range for the cumulative dose and cumulative dose squared states are $0 \leq U_t \leq \min_i\{\frac{c_i}{\gamma_i}\}$ and $0 \leq V_t \leq \min_i\{\frac{c_i\abo}{\gamma_i^2}\}$. For the radiation dose fractions, we allow only multiples of $0.1$ Gy. There is a significant amount of debate, \cite{Brenner2008,KiMeMa08}, as to whether the linear quadratic model shown in equation \eqref{eq:LinearQuadraticExp} applies to large doses. In view of this, we include a constraint that the dose fractions be no more than 5 Gy.

\begin{figure}

\subfloat[Case 1]{\includegraphics[width = 70mm]{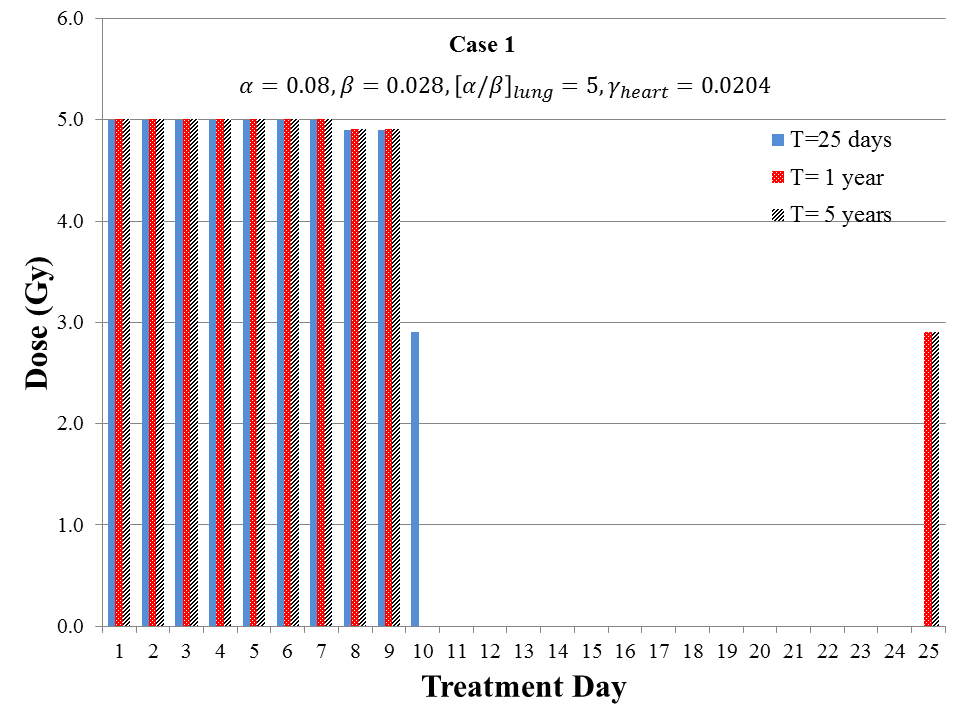}} 
\subfloat[Case 2]{\includegraphics[width = 70mm]{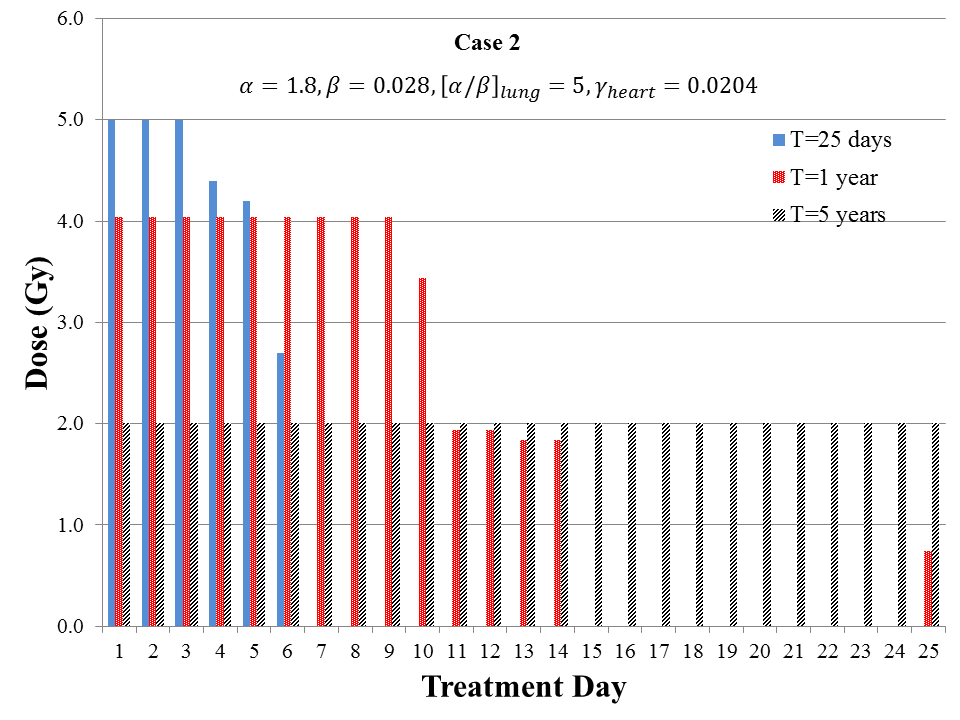}}  
\\
\subfloat[Case 3]{\includegraphics[width = 70mm]{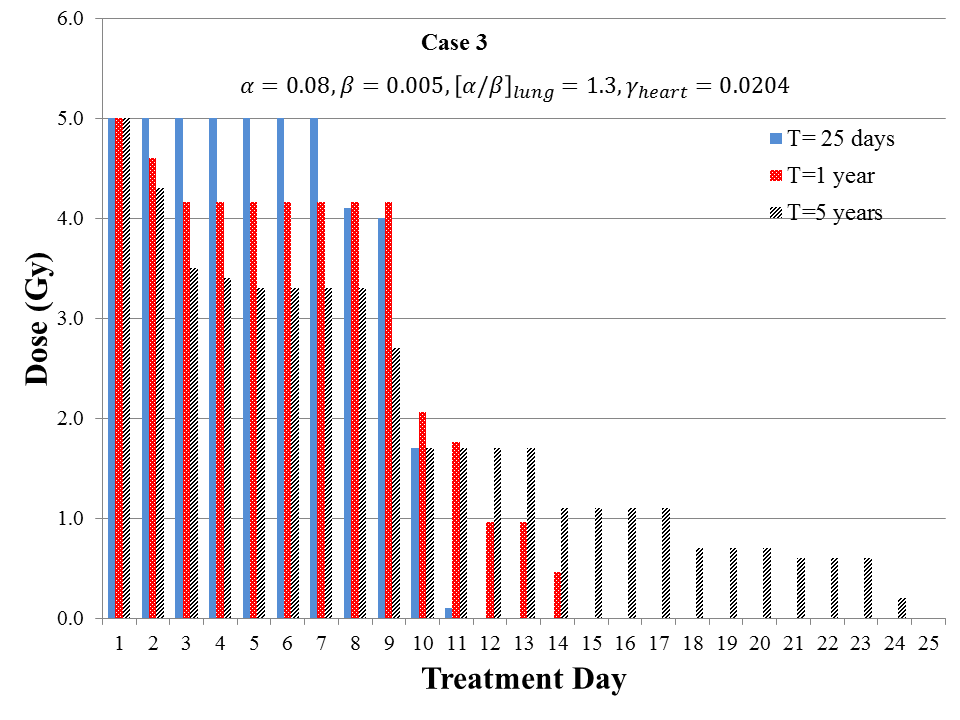}} 
\subfloat[Case 4]{\includegraphics[width = 70mm]{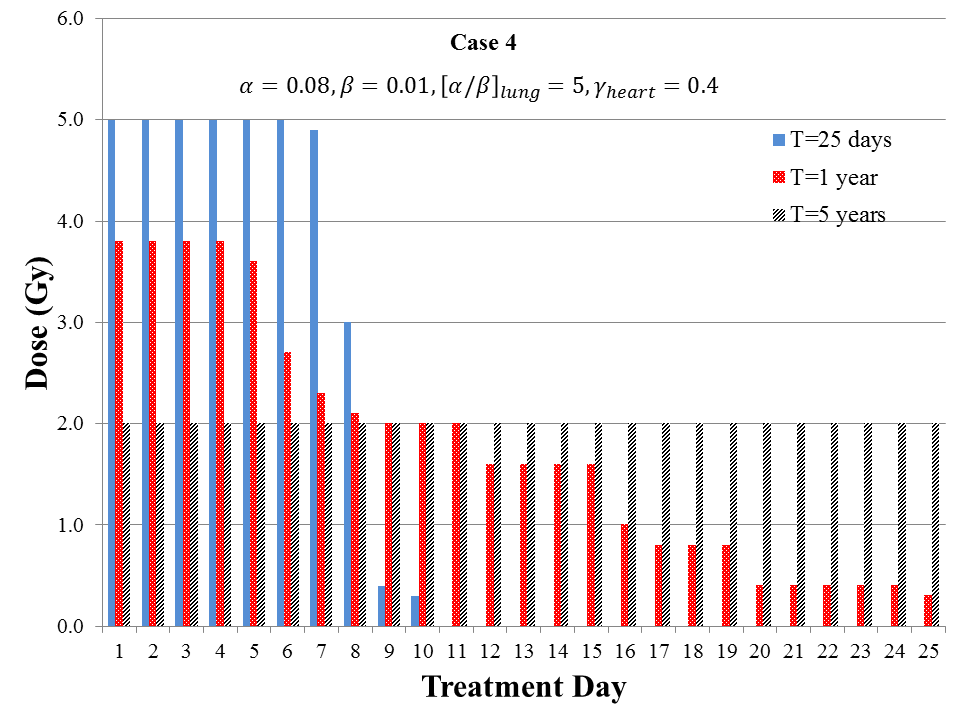}}
\caption{Optimal fractionation schedule with evaluating objective function immediately after treatment, 1 and 5 years after conclusion of therapy. (a) This plot shows that if $\abt\le\min_i\{\abo/\gamma_i\}$ then a hypo-fractionated schedules minimize both tumor cell population at the conclusion of treatment and metastasis risk. In this case we have $\abt=2.86$ and $\min_i\{\abo/\gamma_i\}=60.24$. (b)-(d) In these three cases we have $\abt>\min_i\{\abo/\gamma_i\}$ where an equal-dosage routine minimizes the number of tumor cells at the conclusion of therapy. However we observe that hypo-fractionated schedules are still the best choices when minimizing metastatic risk in short term (see blue and red bars above). Evaluating metastatic risk in long term, optimal schedules change to an equal-dosage routine (gray bars in case 2 and case 4) or a semi-hyper-fractionated schedule (gray bars in case 3), which shows the sensitivity of these structure with respect to $T$, $\alpha$, $\beta$ and OAR parameters. In cases $b$, $c$ and $d$, we have $\abt=64.26, 16, 8 $ and $\min_i\{\abo/\gamma_i\}=60.24, 15.7, 7.5$, respectively.}
\label{fig2}
\end{figure}

Figure \ref{fig2}-(a) shows the optimal fractionation schedule when using $\xi = \frac{1}{3}$. Our model suggests very large initial doses for the first 9 days followed by a smaller dose on the 10th day if $T=25$ days or last day of treatment otherwise. The results are not surprising because previous works  (\cite{MiTaDa12}, \cite{Badri}) have suggested hypo-fractionation in the case where $\abt<\min_i\{\abo/\gamma_i\}$. In order to understand the differences between the optimal schedules in minimizing metastatic risk $T$ years after the beginning of radiotherapy and minimizing tumor population at the conclusion of therapy, we consider various settings for model parameters. We report four of them along with their parameters in Figure \ref{fig2}. We observe that if $\abt\le\min_i\{\abo/\gamma_i\}$ for all $i$, then, similar to the case of minimizing tumor population, a hypo-fractionation schedule is still optimal (case 1 in Figure \ref{fig2}). However if we have $\abt>\min_i\{\abo/\gamma_i\}$ (cases 2, 3 and 4 in Figure \ref{fig2}), depending on the model parameters, a hyper-, semi-hyper (a schedule with structure of a large initial dose that tapers off slowly) or hypo-fractionation schedule is optimal for minimizing metastasis production. Note that in the last three cases standard schedule (25x2 Gy) with equal doses is optimal when minimizing the tumor population at the end of therapy. In addition, we see that for smaller metastatic risk horizon $T$, with the same parameters, the optimal schedule tends to be a hypo-fractionation schedule (see the optimal schedule for case 2, 3 and 4 in Figure \ref{fig2} when $T=25$ days).

\begin{figure}[ht!]
\centering
\includegraphics[width=180mm]{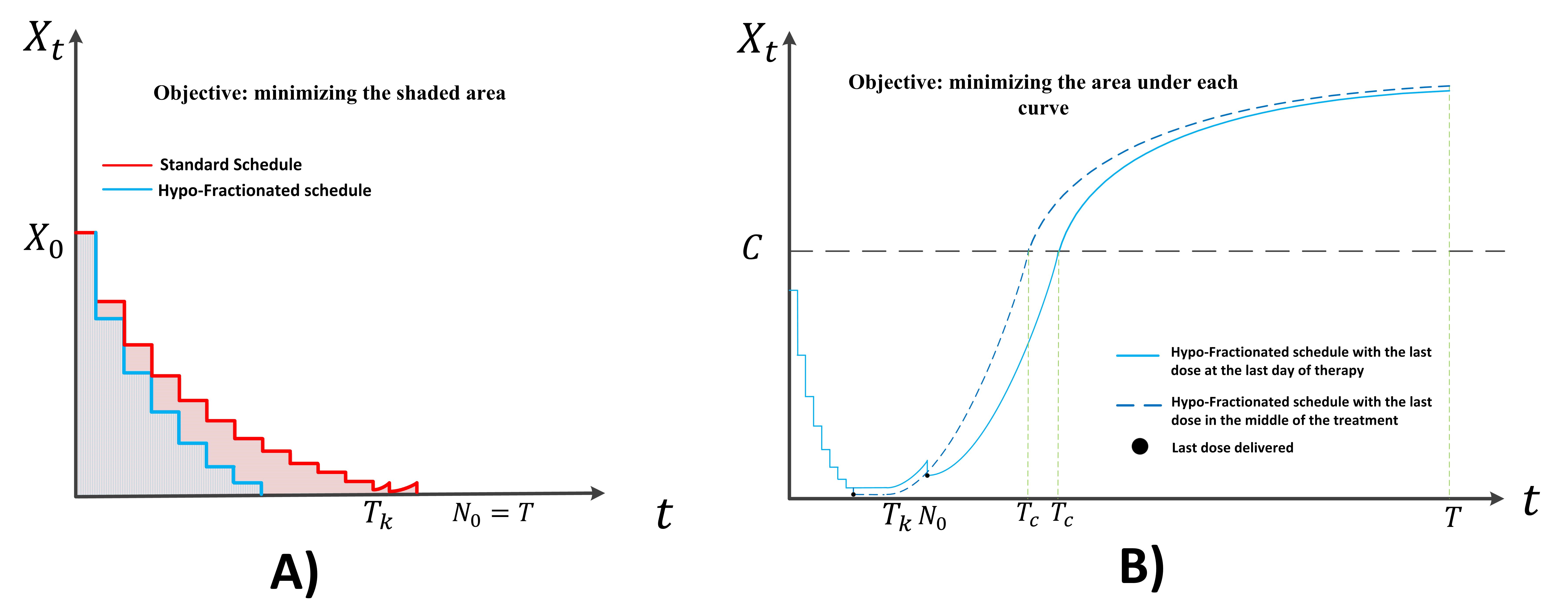}
\caption{Optimality of standard and hypo-fractionated schedules when  A) there are no clonogenic cells at the end of treatment, i.e. full local control. In this case, reducing the tumor size in a short period of time is the optimal strategy. B) insufficient dose is delivered to control the primary tumor and the long term metastatic risk should be considered. In this case in addition to reducing the tumor size at the beginning of the therapy, having small number of clonogenic cells on the last day of treatment is also important. The area under the solid and dashed blue curves explains the property of delivering the last dose on the last day instead of 10th and 15th day in Case 1 and Case 2 of Figure \ref{fig2}, respectively.}
\label{area}
\end{figure}

Figure \ref{area} explains the logic behind optimal schedules described in Figure \ref{fig2}. The objective function described in \eqref{opt} is minimizing $\xi$th root of the area under the tumor size curve, $X_t$. If there are no clonogenic cells at the end of the treatment, i.e. full local control, there is no further risk of metastasis after therapy. In this respect, we need to evaluate the metastasis risk for time interval $[0,N_0]$ ($T=N_0$). Thus the optimal solution is reducing the tumor burden in a very short period of time which supports the optimality of hypo-fractionated schedules (see blue bars in Figure \ref{fig2} and Figure \ref{area}-A). 

However when insufficient dose is delivered to control the primary tumor, the long term metastatic risk should be considered. In this case our optimization problem minimizes $S_1+S_2$ where $S_1$ is the area under the tumor size curve in $[0,N_0]$ and $S_2$ is the area for $(N_0,T]$. Therefore the problem of minimizing tumor burden as quickly as possible (small $S_1$) and maximizing tumor control (which results in small $S_2$) must be optimized simultaneously. If there is a possibility of metastasis creation for a long period of time, we need to evaluate metastasis risk for larger values of $T$, e.g. $T=5$ years, where we have $S_1\ll S_2$. Therefore having minimal number of $X_{N_0}$ is more effective than immediate reduction of the tumor burden. It explains the optimality of standard schedule in Case 2 and Case 4 of Figure \ref{fig2} when we set $T=5$ years (note that in these three cases we have $\abt>\min_i\{[\alpha/\beta]_i/\gamma_i\}$ where an equal dosage regime minimizes $X_{N_0}$). 

Finally we explore an interesting property of the optimal schedules in Figure \ref{fig2}. When we evaluate the long term metastatic risk, if the hypo-fractionated schedule is optimal, we observe that it is preferable to deliver the last dose of radiation on the last day of treatment (see Case 1 and Case 2 in Figure \ref{fig2}). When $T_k<N_0$ and we are unable to fully control the primary tumor, then clonogenic cells start to repopulate during the course of treatment. On the other hand if we deliver the last dose on the last day, a larger population of clonogenic cells are affected which results in smaller $X_{N_0}$ and an improvement in the objective function in long term (compare the solid and dashed blue curves in Figure \ref{area}-B). Note that this result depends on some values of model  parameters, e.g. tumor growth rate, tumor kick-off time and etc, and may not hold for a different tumor with distinct parameters.

\begin{figure}

\subfloat[Sensitivity analysis w.r.t $\xi$]{\includegraphics[width = 85mm]{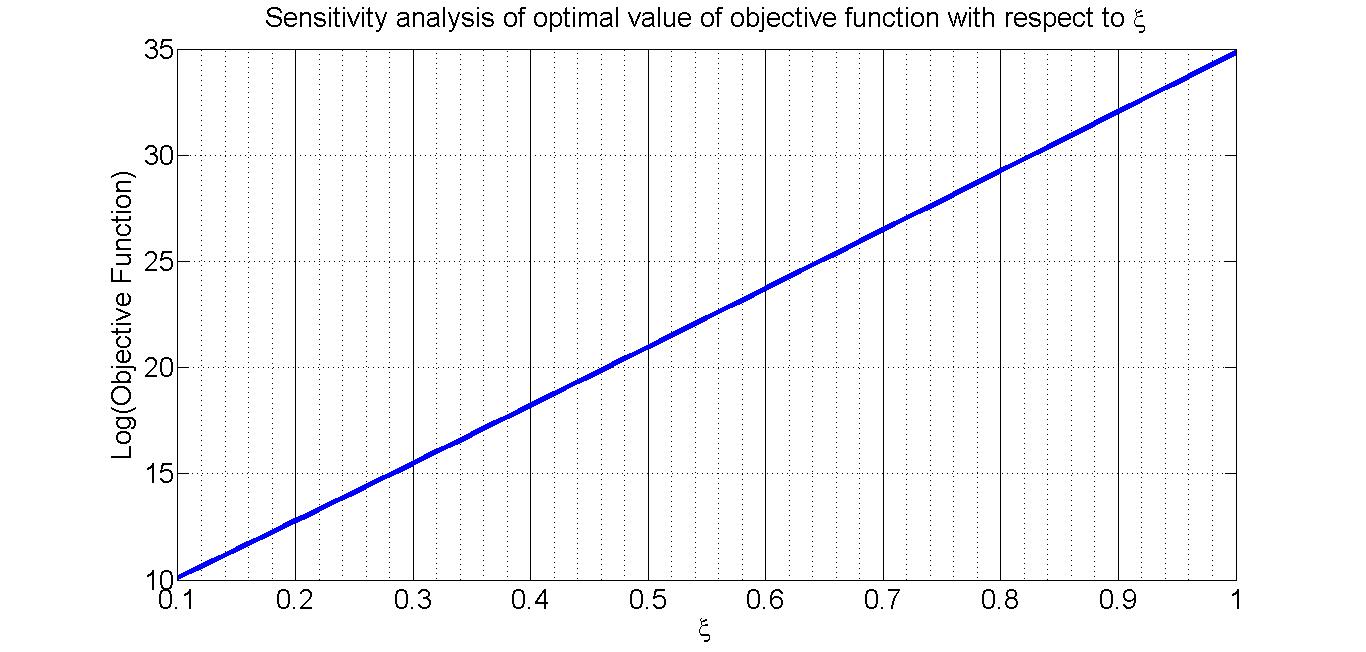}} 
\subfloat[Sensitivity analysis w.r.t $N_0$, $\tau_d$, $T_k$ and $\delta$]{\includegraphics[width = 85mm]{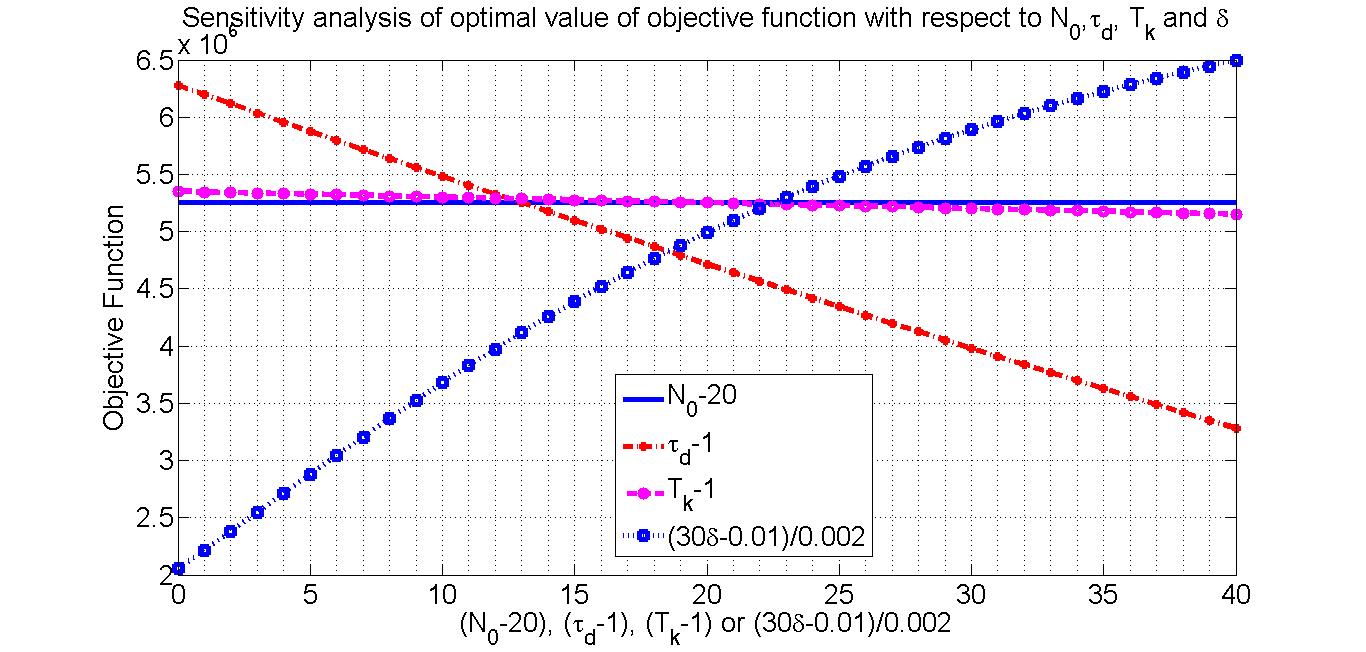}}  

\caption{a) This plot shows the sensitivity of number of metastasis produced by tumor with respect to $\xi$. The structure of the optimal schedule is robust to variation in $\xi$, however the metastasis production is affected dramatically by this parameter. b) This plot shows the sensitivity of metastasis production with respect to $N_0$, $\tau_d$, $T_k$ and $\delta$. The optimal schedules associated with various values of these parameters are the same as the optimal schedule obtained in Figure \ref{fig2}-(a). The effects of treatment duration and tumor kick-off time are negligible in the model's output; however, the exponential and Gompertz growth rate can change the optimal value.}
\label{fig3}
\end{figure}

We perform an extensive sensitivity analysis of the optimal solution with respect to model parameters ($\xi$, $N_0$, $\tau_d$, $T_k$ and $\delta$). We see that after including different parameters the minimization of metastasis risk is still achieved by a hypo-fractionated schedule presented in Figure \ref{fig2}-(a). In addition, we see that for smaller $T$ radiation is concentrated at the beginning of the schedule. Figure \ref{fig3} plots the optimal values of the objective function in \eqref{opt} for different values of model parameters. The structure of the optimal fractionation is relatively robust to model parameters; the only parameters that change this structure are $T$, $\abo$, $\abt$ and $\gamma_i$ (See Figure $\ref{fig2}$). 

We consider the relative effectiveness of an optimized schedule versus a standard schedule (delivering 50 Gy to the tumor with 2 Gy
fractions in 5 weeks \cite{stand1}). In particular, if we denote the approximate metastasis risk under the optimized schedule by $\nu\hat{R}_\xi^{opt}$, and the risk under a standard uniform fractionation by $\nu\hat{R}_\xi^{std}$. Then, the ratio $r_e=(\hat{R}_\xi^{std}-\hat{R}_\xi^{opt})/\hat{R}_\xi^{std}$ will give us a measure of the predicted relative reduction in metastasis risk associated with using the optimized schedule instead of the standard schedule. These results are presented in Table \ref{tab:relative_risk}. In the case of eradicating the primary tumor, we should only consider the metastasis production during the course of treatment ($T=0.07$). In this case, we see a significant reduction in metastasis risk for all values of $\abt$ studied. However if insufficient dose is delivered to the tumor, primary tumor begins to repopulate after therapy and we require to choose a bigger $T$ to evaluate the metastasis risk since cancer metastasizes either during or after treatment ($T=1,2,3,4,5$ years). In general we see that the risk reduction decreases with $\abt$ and the evaluation period, but optimal schedule still provides for a substantial risk reduction for all values of $\abt$ and $T$ studied.

\begin{table}
\begin{tabular}{|c|c|c|c|c|c|c|c|c|c|c|c|c|}
\hline
 & \multicolumn{11}{|c|}{$\abt$} \\
\hline  & 5 & 6& 7& 8& 9& 10& 11 & 12& 13& 14& 15 \\
\hline
$T=0.07$ & $53.16\%$ & $48.97\%$ & $45.55\%$ & $42.74\%$ & $40.39\%$ & $38.40\%$ & $36.70\%$& $35.23\%$& $33.95\%$& $32.82\%$& $31.81\%$ \\
\hline
$T=1$ & $20.93\%$ & $17.10\%$ & $14.39\%$ & $12.37\%$ & $10.82\%$ & $9.58\%$ & $8.57\%$& $7.73\%$& $7.02\%$& $6.42\%$& $5.89\%$ \\
\hline
$T=2$ & $10.72\%$ & $8.76\%$ & $7.37\%$ & $6.33\%$ & $5.52\%$ & $4.87\%$ & $4.35\%$& $3.91\%$& $3.54\%$& $3.22\%$& $2.95\%$ \\
\hline
$T=3$ & $6.94\%$ & $5.66\%$ & $4.76\%$ & $4.08\%$ & $3.55\%$ & $3.13\%$ & $2.79\%$& $2.51\%$& $2.27\%$& $2.06\%$& $1.89\%$ \\
\hline
$T=4$ & $4.88\%$ & $3.98\%$ & $3.34\%$ & $2.86\%$ & $2.49\%$ & $2.2\%$ & $1.96\%$& $1.76\%$& $1.59\%$& $1.45\%$& $1.32\%$ \\
\hline
$T=5$ & $3.63\%$ & $2.96\%$ & $2.49\%$ & $2.13\%$ & $1.86\%$ & $1.64\%$ & $1.46\%$& $1.31\%$& $1.18\%$& $1.08\%$& $0.98\%$ \\
\hline
\end{tabular}
\caption{Metastasis risk reduction of optimized schedule evaluated $T$ years after therapy.}
\label{tab:relative_risk}
\end{table}

\section{Conclusion}
\label{sec:concl}
In this work, we have considered the classic problem of optimal fractionation schedules in the delivery of radiation. We have however done this with the non-traditional goal of minimizing the production of metastasis. This is motivated by the fact that the majority of cancer fatalities are driven by metastasis \cite{GuMa06,weinberg2013biology}, and that this disseminated disease can be very difficult to treat. We addressed this goal by considering the optimal fractionation problem with a novel objective function based on minimizing the total rate of metastasis production, which we argue is equivalent to minimizing the time integrated tumor cell population. We were able to numerically solve this optimization problem with a dynamic programming approach.

We computed radiotherapy fractionation schedules that minimized metastatic risk for a variety of parameter settings. The resulting optimal schedules had an interesting structure that was quite different from what is observed from the traditional optimal fractionation problem where one is interested in minimizing local tumor population at the end of treatment. In the traditional optimal fractionation problem it was observed in \cite{Saberian} that if $\abt\le\min_i\{\abo/\gamma_i\}$ then a hypo-fractionated schedule is optimal and if $\abt>\min_i\{\abo/\gamma_i\}$ then a hyper-fractionated schedule is optimal. In contrast, in the current work we observed that if we are evaluating metastatic risk at the conclusion of therapy then independent of the relationship between $\abt$ and $\min_i\{\abo/\gamma_i\}$, the resulting fractionation schedules for minimizing metastatic risk is a hypo-fractionated structure with large initial doses that taper off quickly. This is due to the structure of the objective function. In order to minimize the time integrated tumor cell population immediately after treatment, it is necessary to quickly reduce the tumor cell population since this is the high point of the tumor cell population over the course of the treatment. If we think of the tumor cell population as quite dangerous due to its metastasis potential, then it is natural to want to reduce their population as quickly as possible.

We observed that the structure of the optimal schedule depends on the length of time for which we evaluated metastatic risk.  In particular if we evaluate metastasis risk in a long time frame (several years) after therapy, and it holds that $\abt>\min_i\{\abo/\gamma_i\}$, then the optimal schedule will have either medium initial doses tapering off slowly (that is, a schedule where $d_1\ge d_2\ge\dots\ge d_{N_0}$) or equal-dosage routine where we have $d_1= d_2=\dots=d_{N_0}$. This is in contrast to when we evaluate metastatic risk over the course of the treatment period, in which case a hypo-fractionated schedule is nearly always optimal. Note that if our evaluation time-frame of metastatic risk is several years after the conclusion of therapy, then we want to minimize the tumor cell population as soon as possible to reduce initial metastatic risk, but we also need to avoid a large tumor population in the long time period after the conclusion of treatment. 

In many of our results, we observed the optimality of hypo-fractionated schedules. Interestingly, previous clinical trials, e.g., the FAST trial for breast cancer \cite{kacprowska2012hypofractionated} and the CHHIP trial for prostate cancer \cite{dearnaley2012conventional}, show the benefit of hypo-fractionation. It should be noted that the motivation for these trials was the low $\abt$ value in prostate and breast tumors. Our results indicate even when the $\abt$ ratio of the tumor is very large, it may still better to deliver a hypo-fractionated schedule when taking metastatic risk into account. The current work provides a possible new motivation for considering hypo-fractionated schedules, i.e., metastasis risk reduction.

In this paper, we have assumed the delivery of single daily fractions and have not considered alternate fractionation schemes such as CHART \cite{HallGiaccia} that deliver multiple fractions a day. Some of our results indicate optimal hypo-fractionation schedules (as obtained for the T=0.07 case). One way to deliver a large amount of dose in a very short period of time is to use such schemes like CHART. However, additional modeling is needed to incorporate incomplete sublethal damage repair due to short inter-fraction time periods.

Another possible interpretation of this work is to view the output $R_\xi$ as the risk of the tumor developing resistance to a chemotherapeutic treatment. This can be achieved by simply viewing the parameter $\nu$ as the rate at which tumor cells develop drug resistance. Due to the severe consequences of drug resistance, this is also an interesting direction for further exploration.

We feel that this type of work minimizing metastatic production opens the potential for a new line of research in the radiation optimization community as well as cancer biology. In particular, there are several important biological phenomena that we have not included. This includes oxygenation status (and history) of cells as well as the vascular structure of the tumor of interest. Lastly, a potentially interesting extension of this work could be to validate our predictions in animal models of metastatic cancer. 

\clearpage
\bibliographystyle{plain}

\end{document}